\documentclass[fleqn,usenatbib]{mnras}

\usepackage{newtxtext,newtxmath}

\usepackage[T1]{fontenc}
\usepackage[utf8]{inputenc}
\usepackage{ae,aecompl}
\usepackage{float}

\usepackage{graphicx}	
\graphicspath{{images/}}
\usepackage{amsmath}	
\usepackage{amssymb}	
\usepackage[usenames]{xcolor}
\usepackage{color}
\usepackage{booktabs}
\usepackage{footnote}
\makesavenoteenv{tabular}
\usepackage[flushleft]{threeparttable}
\usepackage{ulem} 

\title[A wide gap in DS Tau]{Is the gap in the DS Tau disc hiding a planet?}

\author[Veronesi et al.]{
Benedetta Veronesi$^{1,4}$\thanks{E-mail: benedetta.veronesi@unimi.it},
Enrico Ragusa$^{2}$,
Giuseppe Lodato$^{1,4}$,
Hossam Aly$^{1,3}$,
\newauthor
Christophe Pinte$^{4,5}$, 
Daniel~J. Price$^{4}$,
Feng Long$^{6,7},$ 
Gregory J. Herczeg$^{7}$
\newauthor
and Valentin Christiaens$^{4}$
\\
\\
$^{1}$Dipartimento di Fisica, Universit\`a degli Studi di Milano, Via Celoria, 16, Milano, I-20133, Italy\\
$^{2}$School of Physics and Astronomy, University of Leicester, Leicester, United Kingdom\\
$^{3}$Univ Lyon, Univ Claude Bernard Lyon 1, Ens de Lyon, CNRS, Centre de Recherche Astrophysique de Lyon UMR5574,\\ F-69230, Saint-Genis-Laval, France\\
$^{4}$School of Physics and Astronomy, Monash University, Vic 3800, Australia\\
$^{5}$Univ. Grenoble Alpes, CNRS, IPAG, F-38000 Grenoble, France\\
$^{6}$Harvard-Smithsonian Center for Astrophysics, 60 Garden Street, Cambridge, MA 02138, USA \\
$^{7}$Kavli Institute for Astronomy and Astrophysics, Peking University, Yiheyuan 5, Haidian Qu, 100871 Beijing, China\\
}

\date{Accepted XXX. Received YYY; in original form ZZZ}

\pubyear{2020}

\begin{document}
\label{firstpage}
\pagerange{\pageref{firstpage}--\pageref{lastpage}}
\maketitle

\begin{abstract}

Recent mm-wavelength surveys performed with the Atacama Large Millimeter Array (ALMA) have revealed protoplanetary discs characterized by rings and gaps. A possible explanation for the origin of such rings is the tidal interaction with an unseen planetary companion. The protoplanetary disc around DS Tau shows a wide gap in the ALMA observation at 1.3 mm. We construct a hydrodynamical model for the dust continuum observed by ALMA assuming the observed gap is carved by a planet between one and five Jupiter masses. We fit the shape of the radial intensity profile along the disc major axis varying the planet mass, the dust disc mass, and the evolution time of the system. The best fitting model is obtained for a planet with $M_{\rm p}=3.5\,M_{\rm Jup}$ and a disc with $M_{\rm dust}= 9.6\cdot10^{-5}\,M_{\odot}$. Starting from this result, we also compute the expected signature of the planet in the gas kinematics, as traced by CO emission. We find that such a signature (in the form of a `kink' in the channel maps) could be observed by ALMA with a velocity resolution between $0.2-0.5\,\rm{kms}^{-1}$ and a beam size between 30 and 50 mas.

\end{abstract}

\begin{keywords}
 protoplanetary disc --- planet-disc interaction --- dust,extinction --- stars: individual:DSTau --- hydrodynamics --- radiative transfer 
 
\end{keywords}



\section{Introduction}

Recent observations with the Atacama Large Millimeter Array (ALMA) and the SPHERE instrument on the Very Large Telescope (VLT) have revealed protoplanetary discs characterized by sub-structures in their thermal and scattered light emission, including inner holes (e.g. \citealt{dutrey2008,brown2009,andrews11}), gaps, rings (e.g. \citealt{alma-partnership15a,hendler2018,fedele18,dipierro18}), and non-axisymmetric features such as horseshoes (e.g. \citealt{isella2013,zhang14,vandermarel16,canovas16,fedele17, pinilla2017,vandermarel2018,casassus2018,long18,liu2018}), spirals (e.g. \citealt{muto2012,grady2013,garufi13,benisty15,perez2016,stolker16,benisty17}) and shadows (e.g. \citealt{garufi2014,avenhaus2014,benisty17,avenhaus2017}).

The most common substructure in recent ALMA surveys are rings and gaps (e.g \citealt{long18,zhang18} and \citealt{bae18c}). Possible explanations include dust condensation at the snowlines (e.g. \citealt{zhang15}), dead zones (e.g. \citealt{flock15}) and the presence of planets embedded in the disc (e.g. \citealt{dipierro15,dipierro18,bae17a,dong17,rosotti16}). In this work we focus in particular on the planets hypothesis.

Significant information can be extracted from the morphology of gaps. The size and shape of the gap is thought to constraint the mass of the carving planet \citep{kanagawa15b,dong17,dipierro17}, while its position inside the disc and its mass can tell us something about the migration history of the planet inside the disc. 
Planet masses inferred from gap widths are uncertain. First, the time evolution and initial conditions of the disc may lead to different gap shapes and therefore inferred mass for the gap-carving planet. Second, as pointed out also by \cite{pinte19} the gap width alone cannot uniquely constrain the planet mass because changes in the grain density (i.e. the Stokes number) can produce the same gap width with a different planet mass. For these reasons, estimates for planet masses should combine gap widths with other diagnostics.

Recent studies \citep{pinte18,teague18,pinte19,casassus19,pinte20} inferred the location and mass of some planets from the gas kinematics of the discs they were embedded in. Specifically, these studies search in different ways for velocity deviations (localised in both space and velocity) from the unperturbed Keplerian flow of the disc, induced by the presence of a planet, known as ``kinks" and ``Doppler flips".
These revealed giant planets in HD163216 and HD97048.

\begin{figure}
	\includegraphics[scale=0.38]{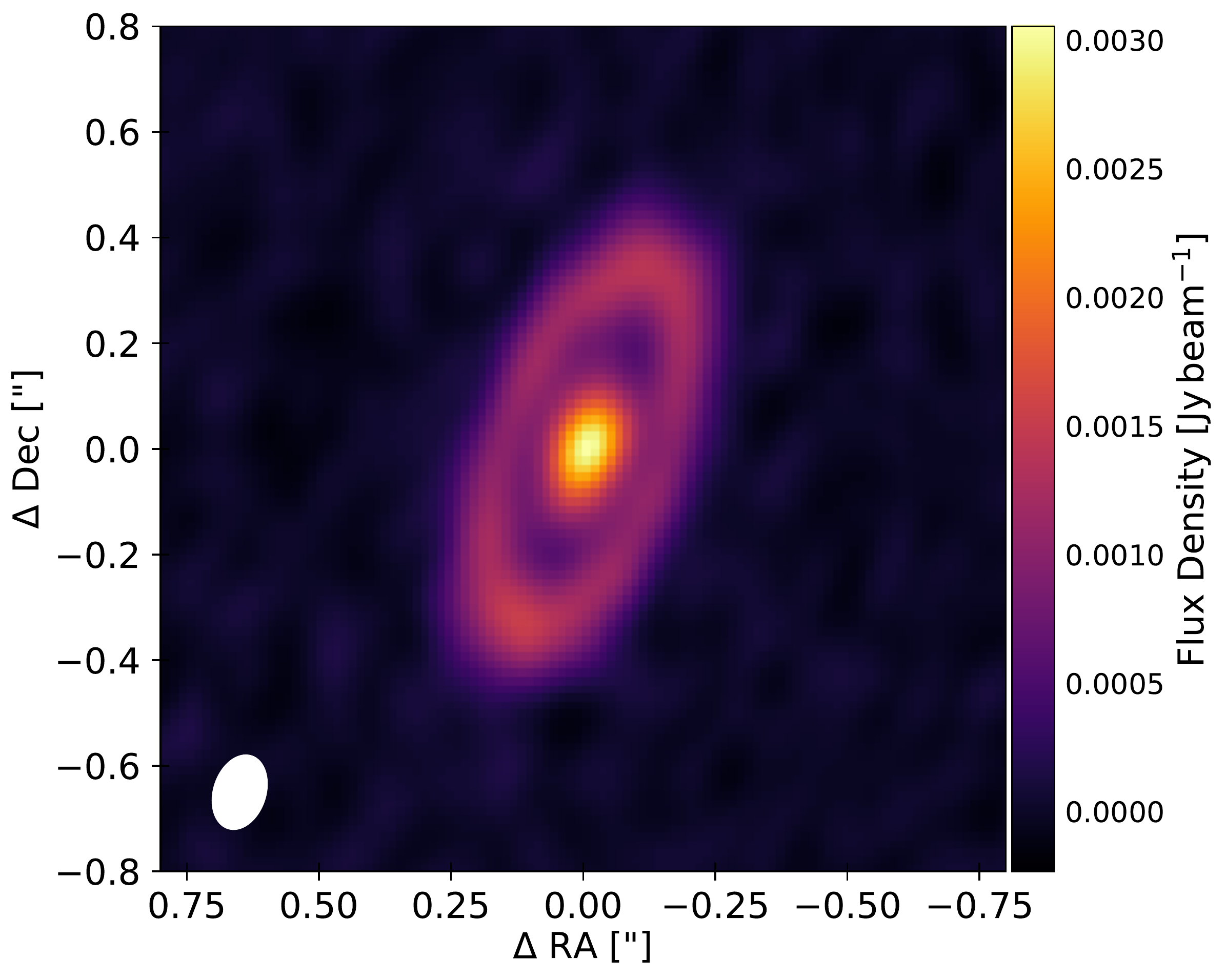}
 	\caption{ALMA observation at 1.3 mm of the system DS Tau, with a beam size of $0.14^{\prime\prime}\times0.1^{\prime\prime}$ wide (image readapted from \citep{long18}).}
    \label{fig:continuumOBS13}
\end{figure}

Here, we study the protoplanetary disc orbiting around DS Tau, an M-type star ($0.83 M_{\odot}$, \citealt{lodato19}) in the Taurus star-forming region located at a distance of 159 pc \citep{gaia18}. It has been observed in ALMA Cycle 4 program (ID: 2016.1.01164.S; PI: Herczeg) in Band 6 at 1.33 mm, at high-spatial resolution ($\sim0.12^{\prime\prime}$, corresponding to $\sim16$ au). This disc shows the widest gap (see Fig.~\ref{fig:continuumOBS13}, \citealt{long18}) of the Taurus survey with a width\footnote{the gap width has been defined as the full width half maximum.} of $27$ au centered at $\approx 33$ au (i.e. $\sim0.2^{\prime\prime}$). 
Assuming the planetary hypothesis for the origin of this gap, \cite{lodato19} estimated that a planet mass of $5.6 M_{\rm J}$ could have carved the gap, under the assumption that for low viscosity discs the gap width scales with the planet Hill radius as $\Delta=kR_{\rm H}$ \citep{dodson-Robinson11,pinilla12,rosotti16, fung16,facchini18}. \cite{lodato19} assumed a proportionality constant $k=5.5$ derived from averaging hydrodynamical simulations results \citep{clarke18,liu2018}. 

In this paper, we present a follow-up study by modeling the continuum emission of the protoplanetary disc orbiting around DS Tau, focusing on the gap observed in the Taurus survey and assuming its origin is due to a planet. By combining 3D hydrodynamical simulations of a suite of dusty protoplanetary disc models hosting one embedded protoplanet with 3D Monte Carlo radiative transfer simulations, we study the gap shape of DS Tau and also analyse the different observational predictions for a planet kink detectable in the gas kinematics. Here, we have chosen to explore the simplest model that can explain the formation of the gap, that is a single planet on a circular, non-inclined orbit, following Occam's razor. Other scenarios could be equally plausible but generally involve a larger number of free parameters. Among these, we might consider: a multiple planet system, a planet on an eccentric \citep{muley19} or inclined orbit, a lower disc viscosity combined with a sub-Neptune mass planet \citep{dong17b,dong18a}. Moreover, a higher resolution image could turn the observed broad disc ring into multiple narrow rings (e.g. HD 169142: \citealt{fedele17} vs \citealt{perez19}), something that at present we cannot exclude but neither explore.

This paper is organised as follows: in Section~\ref{sec:numerics} we describe our numerical method and simulation setup. In Section~\ref{sec:results} we describe the results of the numerical simulations and the fitting procedure we use to find our best model. In Section~\ref{sec:discuss} we discuss our modeling results for the continuum images and we analyse our kink prediction. We finally draw our conclusions in Section~\ref{sec:conclusion}.

\section{Methods}
\label{sec:numerics}

\subsection{Dust and gas numerical simulations}
We perform a suite of 3D Smoothed Particle Hydrodynamics (SPH) simulations of dusty protoplanetary discs, using the code \textsc{phantom} \citep{price18phantom}. 
We adopt the multigrain \citep{hutchison18} one fluid (for St$<1$, \citealt{price15,ballabio18}) method to simulate the dynamics and evolution of dust grains. In this algorithm, SPH particles representing gas and dust are evolved using a set of governing equations describing the gas-dust mixture \citep{laibe14a}. A dust fraction scalar is carried by the particles and is updated according to an evolution equation. This model is only suited for modelling small dust grains (St$<1$) since the formalism lacks the ability to represent large grains velocity dispersion. Moreover, the formalism employs the `terminal velocity approximation' (e.g. \citealt{youdin05}) which greatly simplifies the governing equations and alleviates the need to temporally resolve the dust stopping time, significantly speeding up the computation. Back-reaction from the dust on to the gas is automatically included in this approach. In all our simulations we do not take into account the fluid self-gravity. 

\subsection{Disc models}

\begin{figure*}
	\includegraphics[scale=0.3]{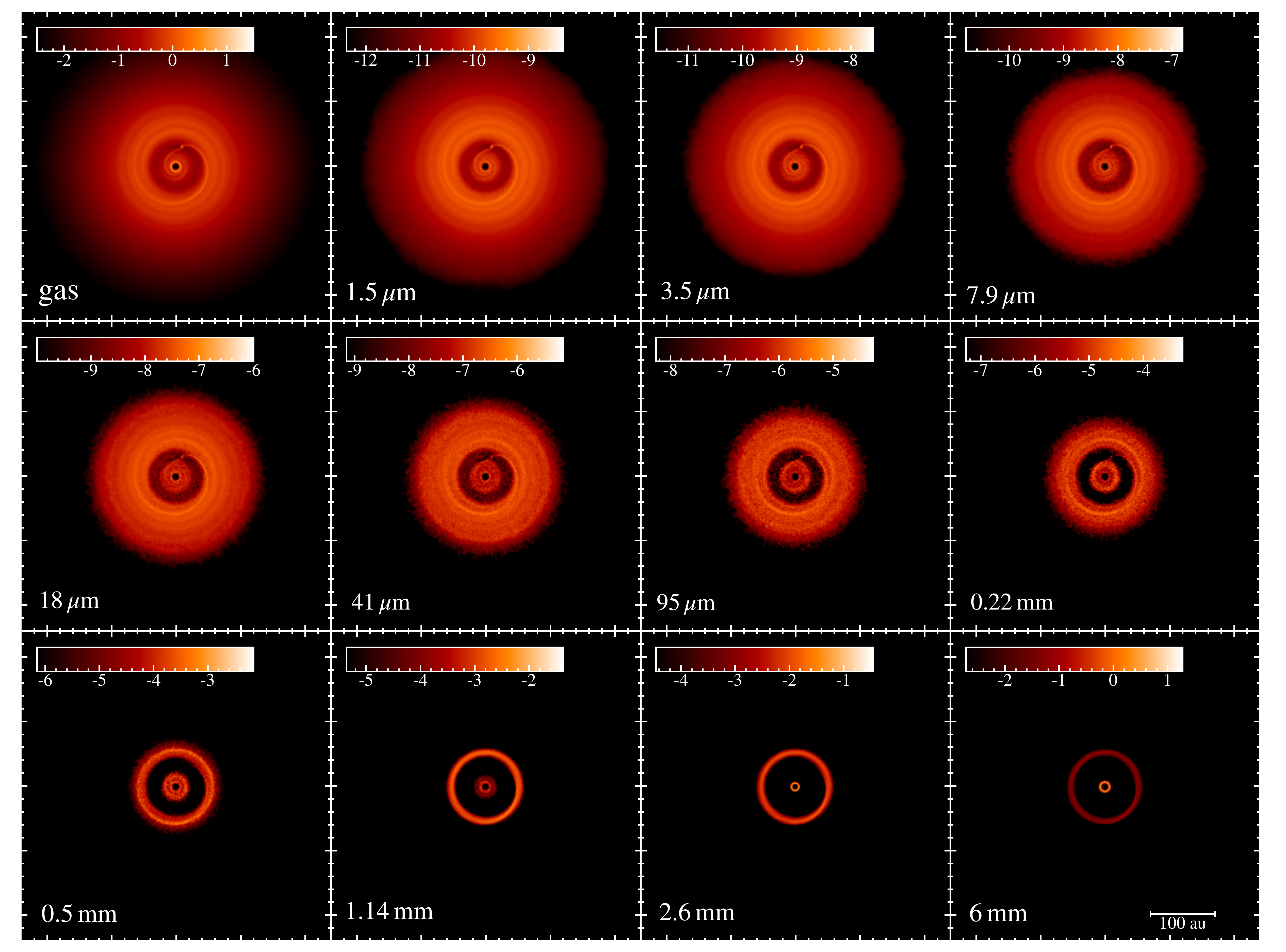}
 	\caption{Rendered images of gas (first panel) and dust surface density (in units of g cm$^{-2}$ on a logarithmic scale) of the $M_{\rm p,0}=2\,M_{\rm{Jup}}$ hydrodynamic simulation after $\approx$140 orbits of the embedded planet at $R_{\rm p,0}=34.5$ au. The dust density is presented for the 11 grain sizes simulated in our model (see Table~\ref{tab:setups}).}
    \label{fig:density_gas}
\end{figure*}

\begin{table}
\caption{Model parameters. $M_\star$ is the star mass as in \citet{lodato19}. $T_{\rm{eff}}$ is the effective temperature we use in our radiative transfer model, assuming the star to be radiating isotropically with a Kurucz spectrum at 3750 K. The disc is located in the Taurus star-forming region at a distance $d$ \citep{gaia18} with an inclination $i$ \citep{long18}. $R_{\rm in}$ and $R_{\rm out}$ are the initial condition for the disc inner and outer radius. $R_c$ and $p$ are respectively the radius of the exponential taper and the power-law index of the gas surface density profile defined in Eq.~\ref{eq:power}. $q$ is the power-law index of the sound speed radial profile (see Eq.~\ref{eq:soundspeed}) and $\alpha_{\rm ss}$ is the effective \citet{shakura73} viscosity. $H/R_{\rm in}$ is the disc aspect ratio at the inner radius. $M_{\rm dust}$ and $M_{\rm gas}$ are the dust and gas disc initial mass. In our model we have $N=11$ dust grains $a_d$, logaritmically spaced between $a_{\rm min}$ and $a_{\rm max}$, with intrinsic grain density $\rho_d$. $M_{\rm p}$ and $R_{\rm p}$ are the mass and the radial position in the disc of the planet.}
\label{tab:setups} 
\begin{center}
\begin{tabular}{lc} 
 \hline
 \hline
Parameters &  Value \\ 
 \hline

$M_\star\, [M_\odot]$ & $0.83$ \\ 
$T_{\rm{eff}}$ [K] & 3800 \\ 
$d\,[{\rm pc}]$ & 159 \\
$i\,[^{\circ}]$ & 65 \\
$R_{\rm in}\, [{\rm au}]$  & $10$ \\
$R_{\rm out}\, [{\rm au}]$ & $100$ \\
$R_{\rm c}\, [{\rm au}]$ & $70$ \\
$p$ & $1.$\\ 
$q$ & $0.25$\\
$\alpha_{\rm SS}$  & $ 0.005$    \\
$H/R_{\rm in}$ & $0.06$ \\
$M_{\rm dust}\, [M_\odot]$ & $ 4.8 \cdot 10^{-5}$ \\
$M_{\rm gas}\, [M_\odot]$ &$4.8 \cdot 10^{-3} $ \\
$ a_{\mathrm{d}}\, [{\rm cm}]$  & $ [a_{\rm min}=1.5\cdot 10^{-4},a_{\rm max}=0.6,N=11]$ \\
$\rho_{\mathrm{d}}\, [{\rm g\,cm}^{-3}]$ &$1$ \\
\hline
$M_{\rm P}\,[M_{\mathrm{j}}]$   & $1,2,2.5,3,5$\\
$R_{\rm P}\,[{\rm au}]$  & $ 34.5 $ \\ 
\hline
\end{tabular}
\end{center}
\end{table}
The parameter choice for our models is motivated by the observations of \citet{long18} (see Table~\ref{tab:setups}).

\subsubsection{Gas and dust}
\label{sec:gasdust_model}
 The system consists of a central star of mass $M_\star=0.83\, M_{\odot}$ \citep{lodato19} surrounded by a gas and dust disc extending from $R_{\rm in}$ = 10 au to $R_{\rm out}$ = 100 au and modelled as a set of $10^{6}$ SPH particles.
The initial gas surface density profile is assumed to be a power law with an exponential taper at large radii, i.e.,
\begin{equation}
\Sigma_{\mathrm{g}}(r)=\Sigma_{\mathrm{c}}\left(\frac{r}{R_{\mathrm{c}}}\right)^{-p} \exp \left[-\left(\frac{r}{R_{\mathrm{c}}}\right)^{2-p}\right]\,,
\label{eq:power}
\end{equation}
where $\Sigma_{\rm c}$ is a normalization constant, chosen in order to match the total disc mass, $R_c=70$ au is the radius of the exponential taper and $p=1$. We adopt a locally isothermal equation of state $P=c_{\rm s}^{2} \rho_{\mathrm{g}}$, with 
\begin{equation}
c_{\mathrm{s}} = c_{\mathrm{s},{\rm in}} \left(\frac{R}{R_{\rm in}} \right)^{-q} ,
\label{eq:soundspeed}
\end{equation}
where $c_{\mathrm{s},{\rm in}}$ is the sound speed at the inner disc radius 
and $\rho_{\mathrm{g}}$ is the gas volume density. We assume $q=0.25$ as the power-law index of the sound speed radial profile. 
The disc is vertically extended by assuming a Gaussian profile for the volume density and ensuring vertical hydrostatic equilibrium 
\begin{equation}
\frac{H}{R} = \frac{c_{\mathrm{s}}}{v_{\mathrm{k}}}=\frac{H}{R_{\rm in}}\left(\frac{R}{R_{\rm in}}\right)^{1/2-q}  ,
\label{eq:aspectratio}
\end{equation}
where $v_{\mathrm{k}}$ is the Keplerian velocity and $H/R_{\rm in}=0.06$ is the aspect ratio at the reference radius $R_{\rm in}$. 
We model the angular momentum transport throughout the disc using the SPH artificial viscosity \citep[see Sec. 2.6]{price18phantom} with $\alpha_{\rm AV}=0.3$, which results in a \citet{shakura73} viscous parameter $\alpha_{\mathrm{SS}} \approx 0.005$.

For the dust we use the same functional form of the initial surface density as for the gas (Eq.~\ref{eq:power}), assuming a dust mass of $M_{\mathrm{dust}}= 4.8\cdot10^{-5}\, M_\odot$. In our SPH simulations the disc mass is fixed, while to fit the continuum emission we rescale it (i.e. the dust mass) as a free parameter (this is possible since the back-reaction is negligible). The dust-to-gas ratio is initially assumed constant for the whole disc extent ($\mathrm{dust/gas}=0.01$), so that the dust has the same vertical structure as the gas. After a few orbits of the planet, the dust settles down forming a layer with thickness \citep{dubrulle95a,fromang09}
\begin{equation}
 H_{\rm d}=H_{\rm g}\sqrt{\alpha_{\rm ss}/({\rm St}+\alpha_{\rm ss})}\,
\end{equation} 
where $\alpha_{\rm ss}$ is the \citet{shakura73} viscous parameter, $H_{\rm g}$ and $H_{\rm d}$ are respectively the gas and dust disc height, St is the Stokes number 
(i.e. $\rm{St}=t_{\rm s}\Omega_{\rm k}$, \citealt{weiden77}).

We perform simulations with the multigrain one-fluid method considering 11 grain sizes logarithmically spaced in a range between 1.5 $\mu$m and 6 mm with a grain size  distribution $dn/da\propto a^{-3.5}$. In this multi-grain method, the dust and gas evolve simultaneously, allowing us to take into account the back-reaction of the dust on the gas, and to simulate different levels of coupling between the two disc components. We note that the back-reaction in our models is in principle negligible for an initial dust-to-gas ratio equal to 0.01. Additionally, even if we consider a later stage in the evolution of our system, the maximum dust-to-gas ratio is still $\sim 0.1$. In Fig.~\ref{fig:density_gas} we show for illustrative purposes the gas (first panel) and the 11 grains dust surface density maps of the disc model with $M_{\rm p,0}=2\,M_{\rm Jup}$, after $\approx140$ orbits of the planet (at the initial planet location $R_{\rm p} = 34.5$ au). Increasing the grain size (from top to bottom) leads to increases in the width and depth of the gap carved by the planet. Moreover, for larger dust grains the dust disc extent is smaller because of radial drift.

\subsubsection{Properties of the embedded planets}
\label{sect:planetprop}
In each disc model we embed one planet with an initial mass of $M_{\rm p}=[1,2,2.5,3,5]\,M_{j}$ at a radial distance from the central star of $R_{\rm p} = 34.5$ au slightly more distant than the 33 au centroid of the gap to account for some migration. The planet orbit is assumed to be initially circular and coplanar. We model the planet and the central star as sink particles, free to migrate and which are able to accrete gas and dust \citep{bate95}. The accretion radius of each planet is chosen to be one quarter of the Hill radius. \cite{lodato19} estimated a planet mass equal to $M_{\rm p}=5.6\,M_{\rm Jup}$ in order to open a gap as wide as the one observed in DS Tau by \cite{long18}. 

We have also attempted to simulate planets with $M_{\rm p}>5\,M_{\rm Jup}$ \citep{lodato19}, but we do not discuss them, since upon an initial analysis the radial flux profile we obtain in these cases is too different with respect to the observations. Moreover, the planet mass derived from the gap width gives an upper limit estimate.

\subsection{Radiative transfer and synthetic observations}
\label{sec:alma}
We compute synthetic observations of our disc models by performing 3D radiative transfer simulations, by means of the \textsc{mcfost} code \citep{pinte06,pinte09}, starting from the results of the hydrodynamical simulations. Our goal is to compute the dust continuum for Band 6 (1.3 mm, \citealt{long18}) and Band 3 (2.9 mm, Long et al. submitted) and the CO, $^{13}$CO and C$^{18}$O isotopologue channel maps. 

The main inputs for the radiative transfer modelling are the gas and dust density structure, a model for the dust opacities and the source of luminosity. We used a Voronoi tesselation where each \textsc{mcfost} cell corresponds to a SPH particle. We adopted the DIANA dust model for the dust opacity \citep{woitke16,min16a}, assuming a fixed dust mixture composed of 70\% silicate, 30\% amorphous carbonaceous. Note that the shape and the width of a gap carved by a fixed planet mass might change for different opacity values and different Stokes number (e.g. for different grain porosity, fluffiness or shape, see \citealt{pinte19}).

The expected emission maps at 1.3 mm and 2.9 mm are computed via ray-tracing, and using $10^8$ photon packets to sample the radiation field, assuming a disc inclination of $i = 65^{\circ}$ \citep{long18}. We use a passively heated model, where the source of radiation is assumed to be the central star, located at the centre of the coordinate system, with parameters described in Table~\ref{tab:setups}. The full-resolution images at 1.3 mm directly produced by \textsc{mcfost} is then convolved with the same Gaussian beam of the observations reported in \cite{long18}, 0.14 $\times$ 0.1 arcsec ($\sim 22 \times 16$ au at 159~pc). 
In order to compute the CO, $^{13}$CO and C$^{18}$O isotopologue channel maps in the J=2-1 transitions we perform radiative transfer simulations, assuming $T_{\rm gas} = T_{\rm dust}$ and that the emission is at LTE. We include in our simulations freeze-out where $T<20$ K and photo-dissociation induced by UV radiation.
We consider a velocity resolution of $0.2\,\rm{kms}^{-1}$. We then convolve the obtained channel maps with a 40 mas beam ($\sim 6.5$ au at the source distance). 

\begin{figure}
		\includegraphics[scale=0.33]{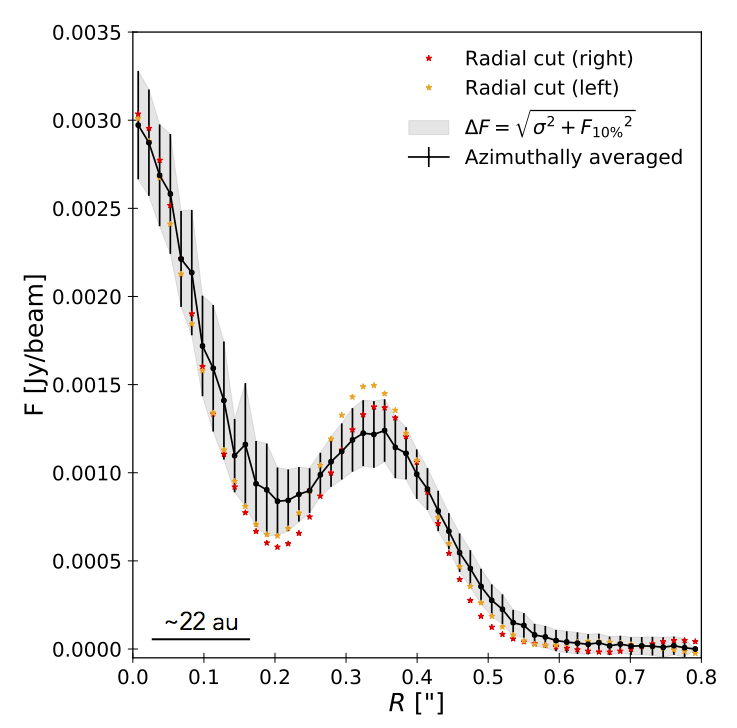}
 	\caption{ Radial flux intensity profile for the ALMA continuum image at 1.3 mm (see Fig.~\ref{fig:continuumOBS13}). In black we show the azimuthally averaged profile while in red and orange the radial cut along the disc major axis. The error is obtained as the root mean square of $\sigma$ and the $10\%$ of the flux value $F_{10\%}=0.1\times F(R)$. In the left corner is reported the projected beam size along the disc major axis $0.14^{\prime\prime}$. 
 	}
    \label{fig:radialOBS13}
\end{figure}

\begin{figure*}
	\includegraphics[scale=0.37]{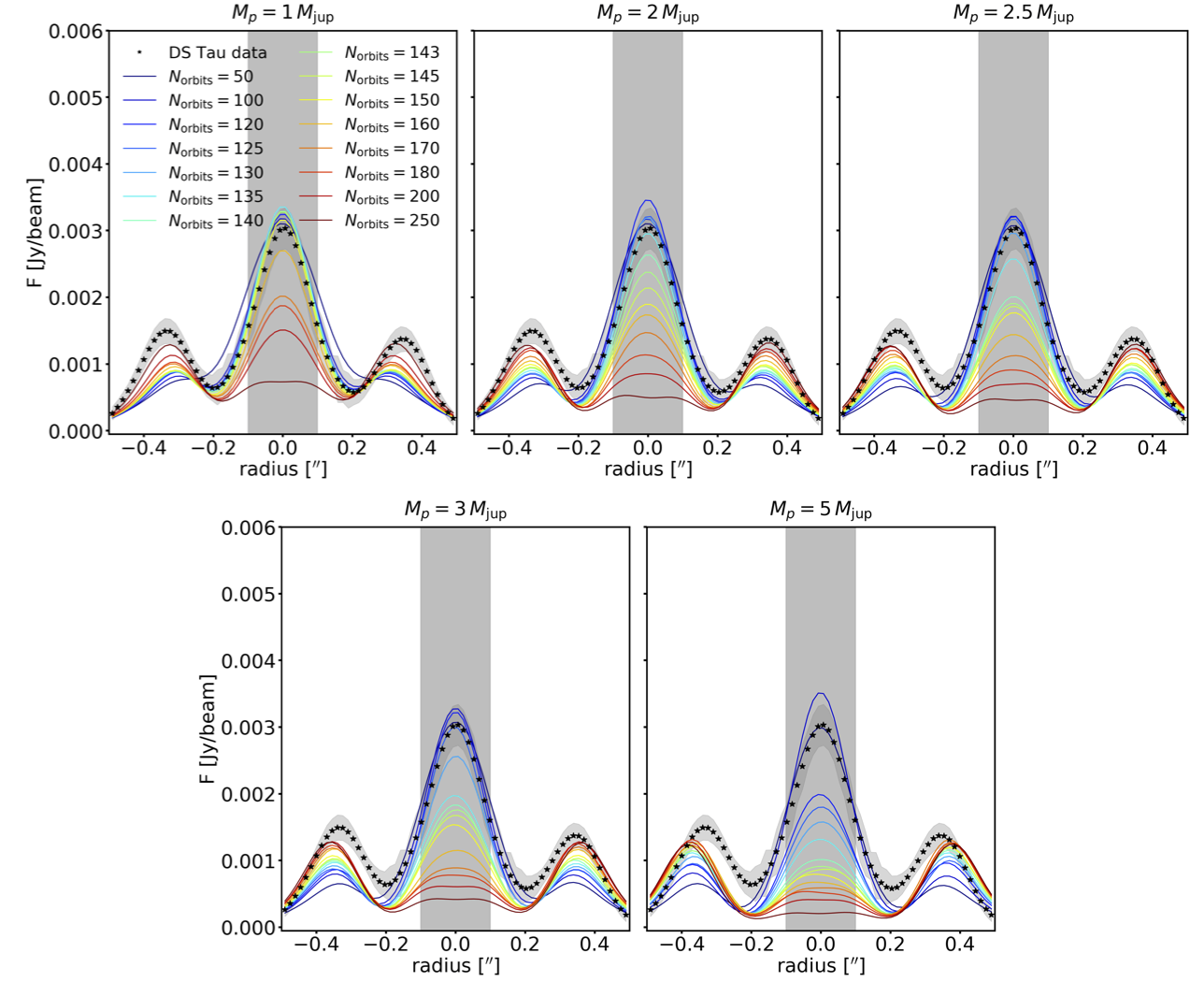}
 	\caption{Radial flux intensity profile along the disc major axis for the models with initial planet masses $M_{\rm p,0}=[1,2,2.5,3,5]\,M_{\rm Jup}$ (from left to right) and for a dust disc mass of $M_{\rm dust}=4.8\cdot10^{-5}\,M_{\rm Jup}$. Different line colours represent the time evolution of the system in terms of planet orbits from 50 (dark blue line) to 250 (dark red line). The grey shaded area $R<0.1"$ corresponds to the region we excluded from the fitting procedure described in Sec.~\ref{sec:innerreg}.
    }
    \label{fig:radialcut_orig}
\end{figure*}

\section{Results}\label{sec:results}

Our main goal is to reproduce the ALMA continuum image at $1.3$ mm observed by \cite{long18}, as shown in Fig.~\ref{fig:continuumOBS13}.
For a more accurate analysis we focus on the radial intensity profile along the disc major axis (i.e. radial cut). Fig.~\ref{fig:radialOBS13} shows both the azimuthally averaged radial intensity profile (black lines) and the radial cut (red and orange markers). We also compute the errors by considering the quadratic sum of the standard deviation $\sigma$ and of the $10\%$ of the flux at each radii \citep{long18}. The standard deviation with respect to the azimuthally averaged radial profile has been computed by deprojecting the disc on a planar surface and binning it into 40 annuli (to match the beam resolution). 
\subsection{Fitting the radial profiles}\label{sec:fit_continuum}

\begin{figure*}
	\includegraphics[scale=0.22]{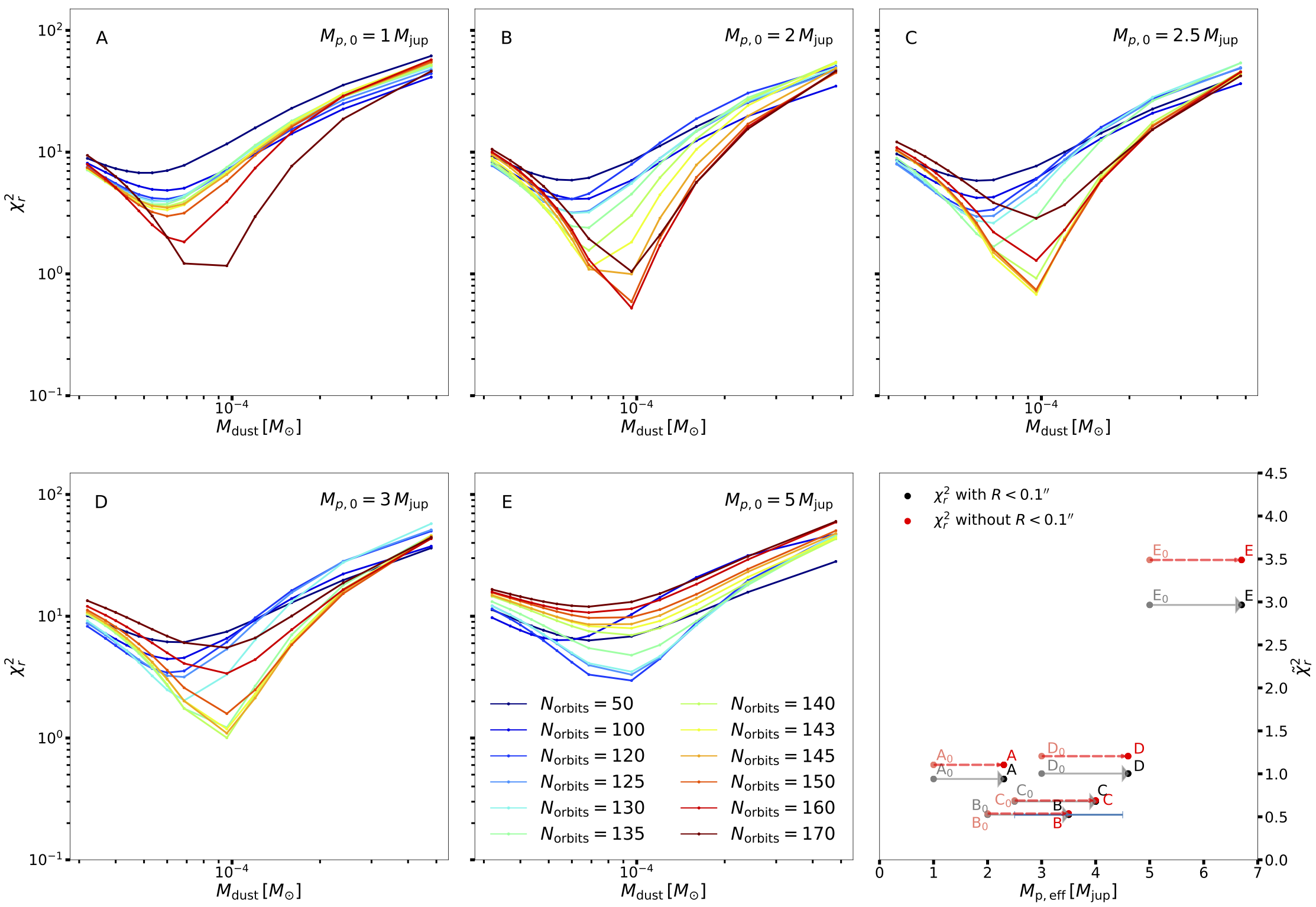}
 	\caption{Reduced chi squared $\chi_r^2$ as a function of the disc dust mass (A,B,C,D,E panels). In each panel we show results obtained for our models with initial planet masses $M_{\rm p,0}=1,2,2.5,3,5\,M_{\rm Jup}$, while different colours correspond to the time evolution in our models, from $N_{\rm orb}=50$ (dark blue line) to $N_{\rm orb}=170$ (dark red line). Minimum reduced chi squared $\tilde{\chi_r}^2=\rm{min}(\chi_r^2)$ (bottom-right panel) from [A,B,C,D,E] panels, as a function of the planet mass in unit of Jupiter masses. The black markers represent the results obtained considering the disc extent $-0.5''<R<0.5''$, while the red ones represent results obtained excluding the inner region $-0.1''<R<0.1''$.} The letters with the $_0$ subscript represent the initial planet mass values, and are connected to the final ones by grey arrows. The value of the planet mass that minimise the $\tilde{\chi_r}^2$ is $M_{\rm p} = 3.5\pm 1\,M_{\rm Jup}$. The error has been obtained via bootstrapping the data.
    
    \label{fig:redchisquare}
\end{figure*}

For a better comparison between our models and observations, and in order to reproduce the correct gap shape, we study the radial intensity profile along the disc major axis. Fig.~\ref{fig:radialcut_orig} shows the radial intensity profile for the disc described in Sec~\ref{sec:gasdust_model} with a dust disc mass of $4.8\cdot10^{-5}\,M_{\rm Jup}$ and an initial planet mass in the range $M_{\rm p,0}=[1,2,2.5,3,5]\,M_{\rm Jup}$. As mentioned in Sec.~\ref{sect:planetprop}, we excluded from our analysis systems with a planet mass $>5\,M_{\rm Jup}$, since their gap shape was too different with respect to the data (as one can already notice in the radial flux intensity profile obtained with $M_{\rm p,0}=5\,M_{\rm Jup}$, see Fig.~\ref{fig:radialcut_orig}). Lines with different colours describe the time evolution of the system (increasing from blue, $N_{\rm orb}=50$, to dark red,  $N_{\rm orb}=250$) while the data are presented as black star markers.  The better match between the models and the observations is found for the $M_{\rm p,0}=2\,M_{\rm Jup}$ case, but the flux needs to be rescaled by a constant factor. In order to fit both the $M_{\rm p}$, the time evolution of the system, the dust disc mass (i.e. the flux intensity), we performed \textsc{mcfost} simulations considering values of gas-to-dust between 10 and 150 for all the orbit snapshots collected in Fig.~\ref{fig:radialcut_orig} (keeping the gas mass constant at $M_{\rm gas}=0.0048$). It is important to highlight that, since the dust-to-gas ratio in our simulation is always $\ll1$ and so the back-reaction of the dust onto the gas is negligible, it is possible to rescale $M_{\rm dust}$, without affecting the disc dynamics. 

As the system evolves, the shape of the intensity profiles shows a similar behaviour for all planet masses. After an annular gap is opened, the inner disc is progressively depleted, both in gas and dust. This is due to a combination of factors. Firstly, the tidal torques produced by the planet reduce the mass accretion rate into the inner portion of the disc (c.f. \citealt{Ragusa16}). Secondly, large dust grains are trapped at the pressure maximum and are thus filtered out from the inner regions \citep{rice06}. Thirdly, there is some spurious evacuation of material from the inner disc, since as the surface density of the gas is reduced, we lose resolution, increasing the SPH artificial viscosity, which in turn speeds up the cavity depletion. Moreover, the increased pressure gradient at the inner boundary accelerates the radial drift in the inner regions. The depletion observed in the inner disc in our models could be explained also by the fact that in our simulation we are not considering dust growth, so we might be neglecting an opacity contribution from larger grains. Finally, we note that by increasing the planet mass, the position of the ring outside the gap moves outward, and that as the system evolves, the planet accretes gas from the disc, so its mass grows with time.

In order to find the best fit model to the data, we thus perform a $\chi^2$ fit assuming as free parameters the planet mass, the scale factor for the dust density and the planet number of orbits. For this last parameter we stop at 170 orbits. After this time, especially for higher planet masses, the inner disc becomes optically thin due to the partially artificial draining of the inner disc. We show the result as a function of the disc dust mass in panels [A,B,C,D,E] of Fig.~\ref{fig:redchisquare}, corresponding to different initial planet masses in the range $M_{\rm p,0}=[1,2,2.5,3,5]\,M_{\rm Jup}$. Different line colours represents different planet orbits in our simulations, from 50 (dark blue line) to 250 (dark red line). We then proceed as follows: for each planet mass we consider the best matching model in terms of minimum $\chi^2$ for the various choices of dust mass and number of orbits. We then compare the best models for different planet masses. In the bottom-right panel of Fig.~\ref{fig:redchisquare} we plot the minimum $\chi^2$ obtained for each planet mass, as a function of the instantaneous planet mass, both with (black markers) and without (red markers, see Sec.~\ref{sec:innerreg}) the inner region $R<0.1''$. Grey (and red) arrows connect the value of the current mass with its initial value. The best fitting model corresponds to the case with an initial planet mass $M_{\rm p,0} = 2 M_{\rm Jup}$, which corresponds to $3.5\pm 1~ M_{\rm Jup}$. We estimated the mass uncertainty via bootstrapping the data. In this way we computed the $\chi^2$ on 10000 new samples taken from the measured data set itself.

\cite{lodato19} estimate a planet mass of the order of 5.6 $M_{\rm Jup}$, assuming that the gap width was $\sim 5.5 R_{\rm Hill}$. Our best fitting model yields a smaller value, which results in a gap width $\Delta/R_{\rm Hill}=7.3$, which is slightly higher compared to the value obtained by \cite{lodato19}. 

\subsubsection{Fitting procedure excluding $R<0.1"$}\label{sec:innerreg}

Due to the possible numerical effects that accelerate the formation of a cavity (discussed in Sec.~\ref{sec:fit_continuum}), and in order to check the robustness of our result, we repeat the fit procedure described in the previous section excluding the flux inside a radius $R<0.1"$ (grey band in Fig.~\ref{fig:radialcut_orig}). This allows us to evaluate the exact portion of the disc which contains the gap and the outer ring, avoiding the possible contamination with regions that could have an evolution influenced by numerical effects. Indeed, we do expect physically to see an inner depletion due to radial drift, on larger timescales with respect to the number of orbits in our models. The results obtained are presented by red markers in the bottom-right panel of Fig.~\ref{fig:redchisquare}. We recover for the planet mass the same value found in Sec.~\ref{sec:results}. We also note that the minimum reduced $\chi^2$ for the $M_{p,0}=1,3,5\,M_{\rm Jup}$ cases is slightly higher than found previously.

\section{Discussion}\label{sec:discuss}

\subsection{Dust continuum} 
In this Section we discuss our results starting from the dust continuum synthetic images at both 1.3 and 2.9 mm wavelengths obtained with \textsc{mcfost}.

\subsubsection{ALMA synthetic images and gap shape}\label{sec:continuum}

\begin{figure*}
	\includegraphics[scale=0.5]{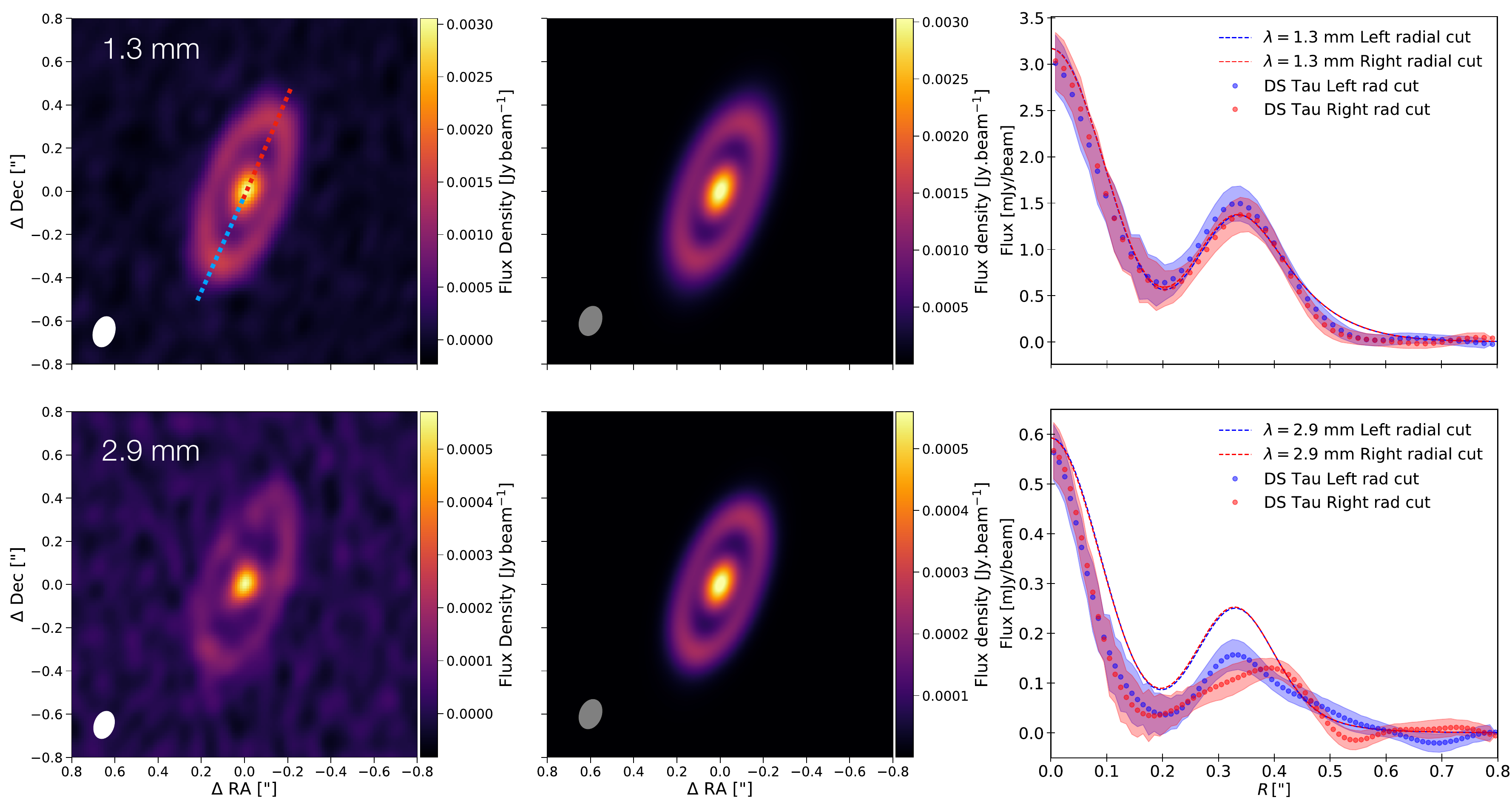}
	\caption{{\it First row}: ALMA observation (left panel) and continuum mock image (center panel) of the DS Tau disc at 1.3 mm. The synthetic image has been computed for our best fit model ($M_{\rm p}=3.5\,M_{\rm Jup}$ and $M_{\rm dust}=9.6\cdot10^{-5}\,M_{\odot}$). The Gaussian beam we take to convolve the full resolution image has been chosen to reproduce the observations reported by \citet{long18} and Long et al. (submitted), respectively $0.14^{\prime\prime}\times0.1^{\prime\prime}$ and $0.13^{\prime\prime}\times0.09^{\prime\prime}$. In the 1.3 mm image we highlight in blue the left side of the major axis and in red the right side. Right panel shows the comparison between the left (blue) - right (red) side of the radial cut along the disc major axis for the data (dot marker) and the modeling (dashed line). {\it Second row}: Same as first row but for the 2.9 mm continuum image. ALMA images have been readapted from \citet{long18} and Long et al. (submitted). }
    \label{fig:best_fit_continuum}
\end{figure*}

The result we obtain from our $\chi^2$ test show that a planet with a mass of $3.5\,M_{\rm Jup}\pm 1$ is the best fit in order to reproduce the gap shape.  The quoted uncertainty clearly refers to the specific model that we assume and additional systematic uncertainty arises if one considers extra degrees of freedom (e.g. multiple planets, orbit eccentricity and inclination, lower viscosity). 
Moreover, having kept as free parameters the dust disc mass and the evolution time of our system, we also obtain that the disc should have a gas-to-dust ratio of 50 with a dust mass of $9.6\cdot10^{-5}\,M_{\odot}$ (similar to what we have fixed in our SPH simulations). 

Concerning the time evolution of the system, the best outcome for our modeling is reached in different time for different planet masses. This is to be expected, since more massive planets carve their gaps faster compared to smaller ones. We also note that by the end of our simulations the system has not reached a steady state configuration. However, it is important to remember that in reality such a quasi steady-state condition does not exist: the disc and planet secularly evolve due to planet migration and accretion processes (note that, often, modeling efforts \emph{assume} that the planet does not migrate nor accrete, and that is why they are able to reach a quasi-steady state: this is not the approach that we have adopted here). What is important is that the number of orbits in our models is enough to avoid transients on a short time scale and that the initial conditions have washed out. This is certainly achieved in our case, as we evolve the simulation for $\gtrsim$ 100 orbits.

Fig.~\ref{fig:best_fit_continuum} shows the comparison between the data and our modeling for the DS Tau disc for both 1.3 (first row) and 2.9 (second row) mm wavelengths. Left panels show the ALMA continuum images from \cite{long18} for the 1.3 mm wavelength and from Long et al. (submitted) for the 2.9 mm one. Center panels show the synthetic images, that have been computed for our best fit model (performed on the 1.3 mm case) parameters: $M_{\rm p}=3.5\,M_{\rm Jup}$, $M_{\rm dust}=9.6\cdot10^{-5}\,M_{\odot}$ and $t=145$ orbits (discussed in Sec.~\ref{sec:fit_continuum}). The right panels of Fig.~\ref{fig:best_fit_continuum} compare radial cuts of the flux intensity profile along the disc major axis obtained from the synthetic image at 1.3 and 2.9 mm (top and bottom center panel of Fig.~\ref{fig:best_fit_continuum}) with the ALMA data (top and bottom left panel). The left (blue) and right (red) side of the radial cut correspond to the left and right side of the disc major axis. Dot markers represent the data, while the models are in dashed lines.

By looking at the synthetic images, the gap shape and the flux of the observation are recovered in our models at 1.3 mm and 2.9 mm. We note that in both cases we end up with a slightly higher peak flux with respect to the observation. An interesting point is that by looking at the real ALMA image at 1.3 mm the ring has a brighter spot in the lower part. This can be due to an inclination effect, though this feature is missing in our model. Also, if we consider the 2.9 mm comparison, the data show an asymmetry in the ring that seems to be more elongated in the right-end side of the major axis. This feature is also not recovered in our model. By a comparison of the synthetic images obtained for the two wavelengths, the 2.9 mm one seems more compact with respect to the other one due to radial drift (see also Fig.~\ref{fig:density_gas}). We will discuss in Sec.~\ref{sec:dusttrapp} what we can learn from this modeling regarding the dust distribution inside the disc. 

The peak of our radial profiles at both wavelengths is slightly higher with respect to the data. This can be due to different dust opacity or to the fact that in our fitting procedure we excluded the flux contribution from a region with radius $R<0.1^{\prime\prime}$. If we consider the 1.3 mm case, for which we have reduced $\chi^2$ fitting process, the gap shape of our model reproduces the one of the data. Instead, for the 2.9 mm case, we note that the model flux is higher by a factor of 0.7 with respect to the data for the entire disc extent.

\begin{figure*}
	\includegraphics[scale=0.5]{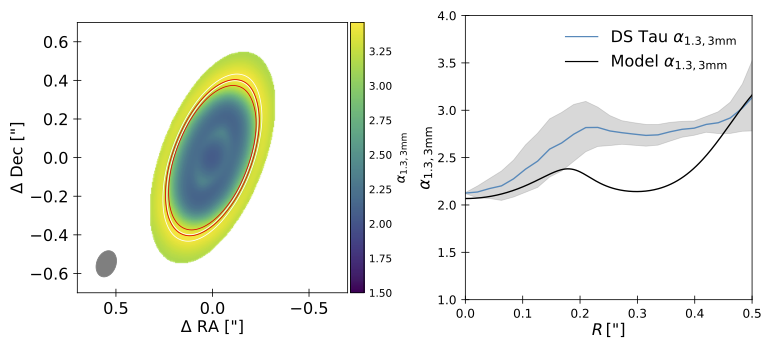}
	\includegraphics[scale=0.3]{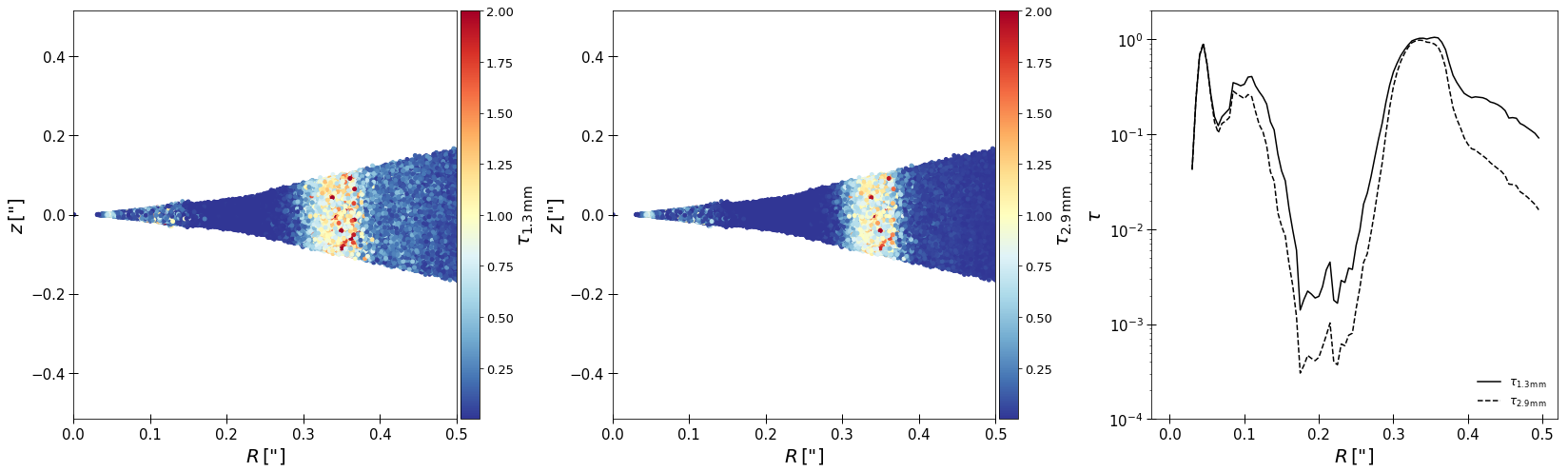}
	
 	\caption{\textit{First row}. Map (left panel) and radial cut along disc major axis (right panel) of the spectral index $\alpha_{1.3,2.9 \rm{mm}}$. Red and white contour lines in the left panel indicate the approximate size of the disc in the 1.3 (white line) and 2.9 (red line) mm images. Central panel shows the spectral index $\alpha$ obtained by Long et al. (submitted). \textit{Second row}. Left and central panels: vertical section of the optical depth as a function of the radius at 1.3 mm and 2.9 mm, respectively. Right panel: azimuthal and vertical average of the optical depth as a function of radius. The solid line represents $\tau$ at 1.3 mm while the dashed line at 2.9 mm. The disc appears to be generally optically thin, but approaches $\tau=1$ in the ring.
    }
    \label{fig:spectral_alpha}
\end{figure*}
\subsubsection{Dust trapping?}\label{sec:dusttrapp}
Having multiwavelength data for the same system allows to analyse the dust distribution inside the disc. In particular, a change in the spectral index $\alpha$ provides information about the grain size in discs \citep{testi14a}. When dust trapping occurs, $\alpha$ is expected to be $\approx 2$, which is smaller than the value tipically found in the ISM, $\alpha_{\rm ISM}\simeq 3.5{\rm -}4.0$ \citep{draine06,ricci10}. Our simulations do produce dust traps (see Fig. \ref{fig:density_gas}), such that larger grains collect in narrower rings. We can thus test whether such traps also show up as a change in the spectral index.

\begin{figure*}
	\includegraphics[scale=0.33]{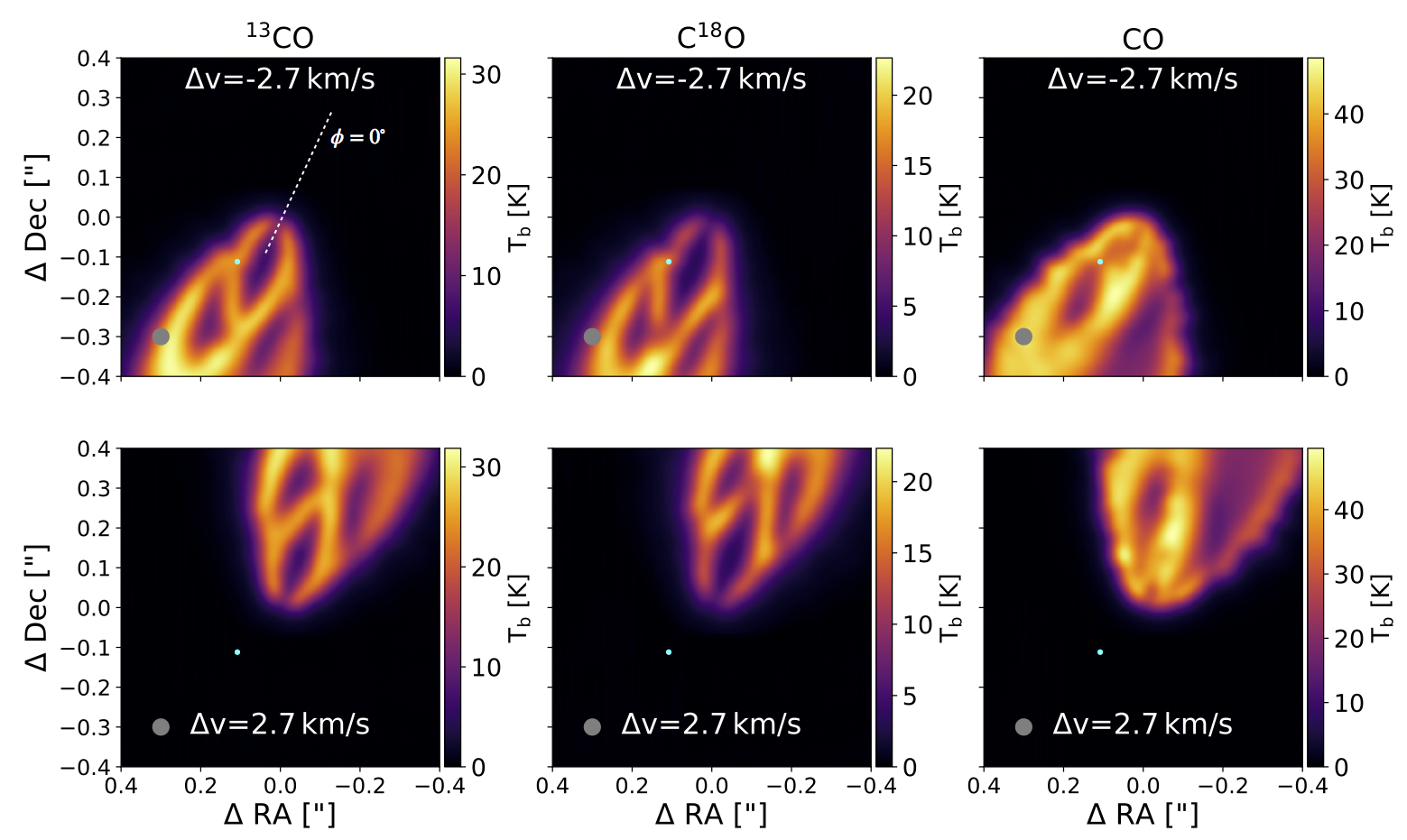}
 	\caption{Synthetic $^{13}$CO, C$^{18}$O and CO ALMA channel maps (J=2-1 transitions) for our best model with $M_{\rm p}=3.5\,M_{\rm Jup}$ (cyan dot). The disc inclination is $i=65^{\circ}$ \citep{long18}, the images have been convolved with a 40 mas Gaussian beam. The azimuthal planet position is $\phi=225^{\circ}$ with respect to the reference case ($\phi=0^{\circ}$, dashed white line).  We notice that the kink produced by the planet is detectable at $-2.7$ km/s with a velocity resolution of $0.2\,{\rm km s}^{-1}$. 
    }
    \label{fig:channel_225azimuth_vres05_40mas}
\end{figure*}

Since we obtained both the 1.3 and 2.9 mm synthetic images, we compute the spectral index $\alpha_{1.3,2.9 \rm{mm}}$ 

\begin{equation}
    \alpha_{\nu_{1},\nu_{2}}=\frac{\log(\nu_2 F_{\nu_2}) - \log(\nu_1 F_{\nu_1}) }{\log\nu_2 -\log\nu_1}\,,
\end{equation}

with $\nu_1 = 1.3$ mm and $\nu_2 = 2.9$ mm. 
In Fig.~\ref{fig:spectral_alpha} we show the resulting spectral index, both as a map (left panel) and as a radial cut along the disc major axis (central panel) in comparison with the $\alpha_{1.3,2.9{\rm mm}}$ radial profile (blue line) found by Long et al. (submitted). We note that at $\simeq0.3^{\prime\prime}$ we have a spectral index $\alpha_{1.3,2.9 \rm{mm}}\simeq2$.
This can be due to two different reasons: there could be dust trapping in the ring region, or the disc may be optically thick. Thus, we compute with \textsc{mcfost} the disc optical depth $\tau_\nu$, where in each cell $\tau_\nu$ is computed from the center of the cell to  $z=+\infty$ (and $-	\infty$). Then, for each Voronoi cell of the model we sum the optical depth towards the $+z$ and the $-z$ direction. The results are displayed in the lower row of Fig.~\ref{fig:spectral_alpha}. In the first two panels we show the vertical cut of the map of the optical depth for each cell. In these plots \textsc{sph} noise is visible, so we also compute the azimuthal and vertical average of the optical depth within concentric annuli. The results of the azimuthally and vertically averaged optical depths are shown in the right panel of Fig.~\ref{fig:spectral_alpha}, with a solid line for 1.3 mm and a dashed line for 2.9 mm. While the disc appears to be generally optically thin, the ring is marginally optically thick. Thus, we cannot draw any firm conclusion on the origin of the spectral slope. Comparing our results with the one presented in Long et al. (submitted), the shape of the spectral index radial profile is similar between the model and the data. Indeed, we qualitatively reproduce the features observed in the data: a local minimum and maximum, with an increase in spectral index at larger radius. 
The actual value of the spectral index is different between the two cases. This can be due to dust grain properties (maximum grain size, dust composition, morphology) in our model not in agreement with the real conditions of DS Tau. This results in different spectral index and different opacity in the optically thin case. Moreover, we point out that considering higher levels of porosity and fluffyness for the dust produces differences in the spectral index profile with respect to the compact case \citep{kataoka14a}.

However, this observed decreasing of $\alpha$ could also be motivated if we consider the presence of larger grains near the ring region. In general, a more precise estimate of the optical thickness of the disc is necessary in order to confirm it.

\subsection{Is it possible to detect the planet from gas kinematic?}\label{sec:kink}

The ubiquity of rings and gaps in recent observations of proto-planetary discs poses one important question: if all these structures have been formed by embedded planets, why are we not able to observe them, apart from few cases such as PDS 70 \citep{keppler18,isella19,mesa19}? A possible way of answering this question and understanding what is the correct origin scenarios for these structures is to study the kinematics of these systems, in particular looking for kinks. 

Assuming that the observed gap is due to a planet with a mass of $M_{\rm p}=3.5\,M_{\rm Jup}$ (see Sec~\ref{sec:fit_continuum}), we check if it can be detected via ``kinks" in the gas channel maps. Therefore, we study the kinematics to determine whether the observational capabilities of ALMA would allow us to detect such a planet at such a distance ($\approx 30$ au) from the central star. For our purpose we choose a channel velocity resolution of 0.5 km/s and an angular resolution of 40 mas. Fig.~\ref{fig:channel_225azimuth_vres05_40mas} shows the computed channel maps for the three CO-isotopologue, for two different channels, with $\Delta$v equals to -2.70 and 2.70 km/s from the systemic velocity. We assure the planet is at an azimuthal position of $\phi=225^{\circ}$ with respect to the reference case (dashed white line). The planet position is plotted in cyan. We note that a kink at the planet location is visible in the $\Delta v = -2.7$ km/s channel but not in the symmetrical one at $\Delta v = 2.7$ km/s. 

In Fig.~\ref{fig:ALLchannels} and Fig.~\ref{fig:ALLchannels2} in the Appendix, we collect channels from $-4.10$ to $-1.30$ km/s and from $1.30$ to $4.10$ km/s. The kink appears localised both in space and in velocity in channels from -3.5 to -1.9 km/s. It is necessary to highlight that there are other similar features, e.g. in channels from 2.10 to 2.70 km/s, which could actually be due to the inside borders of the gap. As a first analysis, the main difference between the two features is that for the planet-induced kink there is an asymmetry between the left and right side of the disc, while for the geometric feature the deviation is symmetric.

To check what is the threshold needed in (spatial and velocity) resolution to be able to observe it, we perform a further analysis, varying the azimuthal angle, the angular and the velocity resolution. We also performed a test in order to determine what is the minimum mass detectable in this system according to our models ($M_{\rm p,fin}=2.3\,M_{\rm Jup}$), at a distance of $\approx 30$ au from the star. The results of this test are presented in Appendix~\ref{app:1mjup}.
\begin{figure*}
	\includegraphics[scale=0.33]{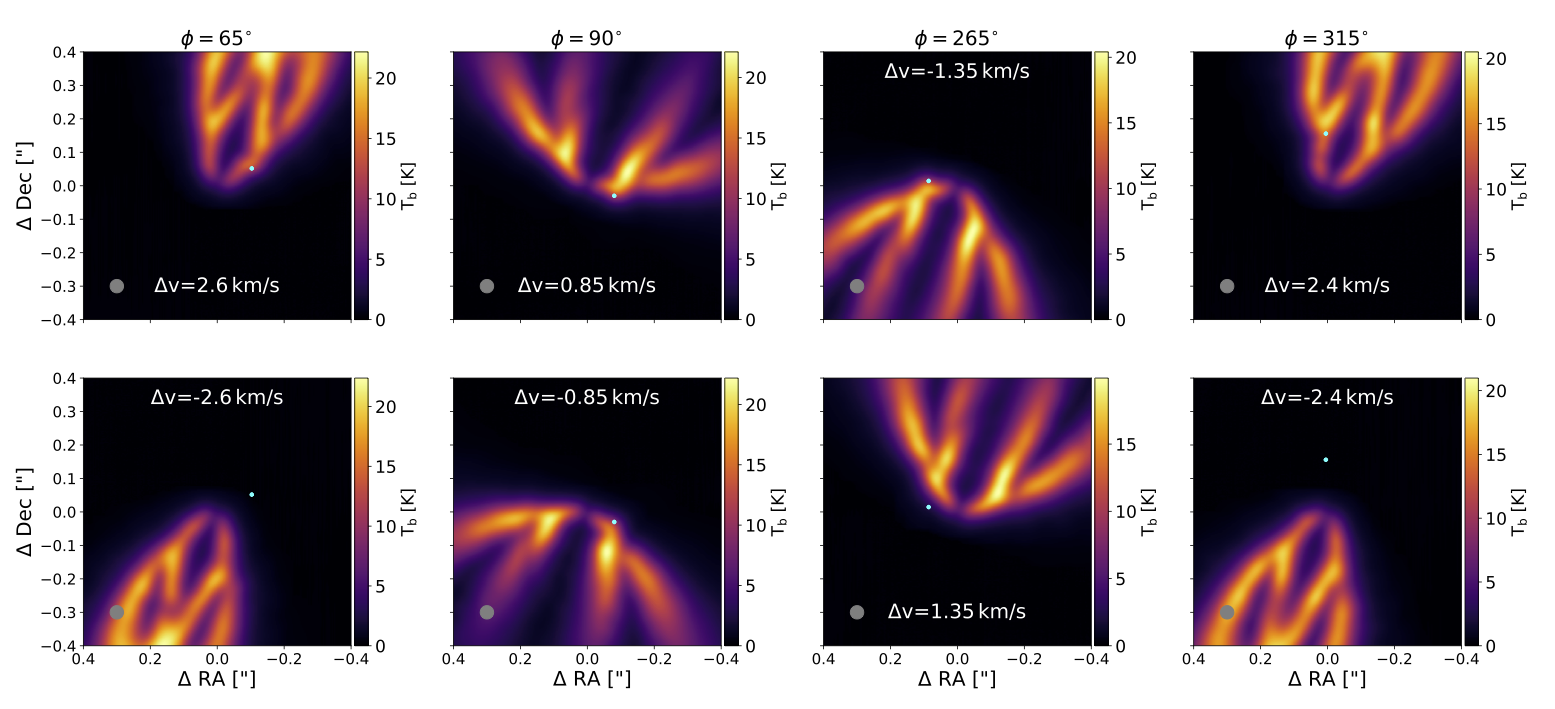}
 	\caption{Synthetic C$^{18}$O ALMA channel maps (J=2-1 transitions) for our best model with $M_{\rm p}=3.5\,M_{\rm Jup}$ (cyan dot). With respect to Fig.~\ref{fig:channel_225azimuth_vres05_40mas} we changed the planet azimuthal position, $\phi=[65^{\circ},90^{\circ},265^{\circ},315^{\circ}]$ (from left to right).
    }
    \label{fig:channel_DIFF_azimuth1}
\end{figure*}

\begin{figure*}
 	\includegraphics[scale=0.33,trim={4.5cm 2.cm 2.cm 2.cm}, clip]{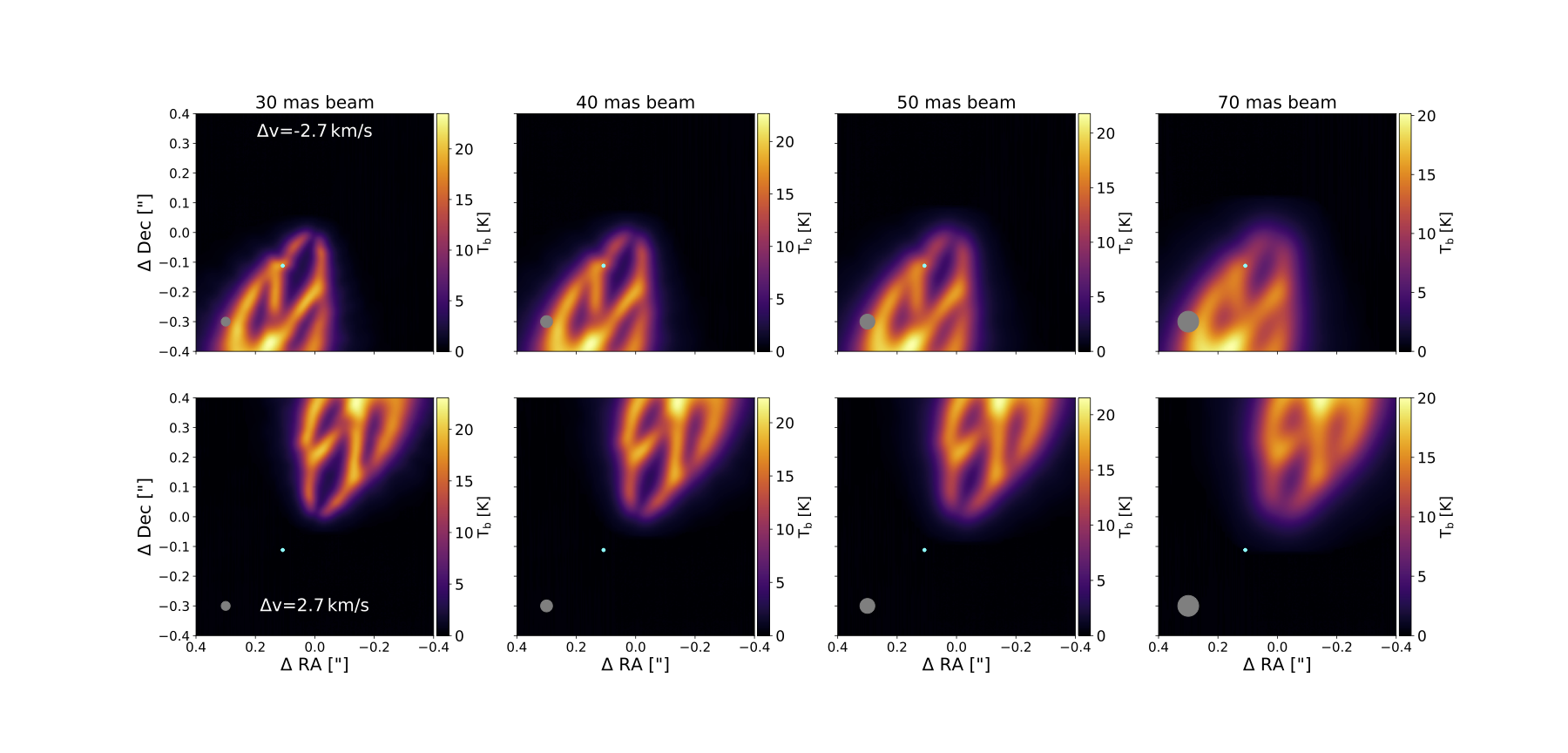}  
 	\caption{Channel maps for C$^{18}$O isotopologue with a velocity resolution of 0.2 km/s and a beam size of [30,40,50,70] mas (from left to right). The azimuthal position of the planet is $\phi=225^{\circ}$. 
    }
    \label{fig:channel_DIFF_beam}
\end{figure*}
\subsubsection{Changing the azimuthal angle}
 
To be able to judge the robustness of our prediction of a velocity kink feature shown in Sec.~\ref{sec:kink}, we need to investigate the dependence of this feature on the azimuthal position of the planet, the velocity resolution, and the beam size. Fig.~\ref{fig:channel_DIFF_azimuth1} shows the channel maps for the C$^{18}$O isotopologue (as best case), varying the azimuthal planet position ($\phi=65^{\circ},90^{\circ},265^{\circ},315^{\circ}$ from left to right). For each azimuthal position, the first row shows the channel in which the planet kink is detectable at different $\Delta$v with respect to the systemic one, while the second one is its symmetrical one. 
Taking into account also the channel maps obtained with $\phi=225^{\circ}$ in Fig.~\ref{fig:channel_225azimuth_vres05_40mas}, we highlight that the kink feature changes for different planet azimuth. Also it appears more detectable for the $225^{\circ}$, $265^{\circ}$ and $315^{\circ}$ cases. It is interesting to note that when the planet is closer to the disc minor axis ($\phi=90^{\circ}$ or $\phi=265^{\circ}$) the kink appears to be visible in a channel and in its symmetrical one.

\subsubsection{Changing the angular resolution}
We also computed channel maps for different beam size and channel velocity resolution, keeping the planet azimuth fixed at $\phi=225$. Fig.~\ref{fig:channel_DIFF_beam} shows the results obtained with beam size of $30,40,50,70$ mas.  The velocity resolution is 0.2 km/s.  Reaching a beam resolution of both 30,40 mas would allow to clearly detect the kink at the planet location, while at 50 mas it is barely visible and at 70 mas it is not detectable.

\subsubsection{Changing the velocity resolution}
\begin{figure*}
 	\includegraphics[scale=0.6]{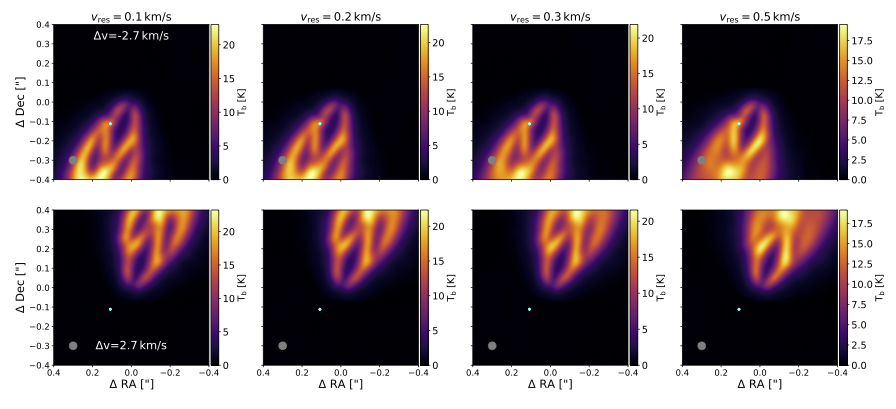}
 	\caption{Channel maps for C$^{18}$O isotopologue with a velocity resolution of 0.1, 0.2, 0.3 and 0.5 km/s (from left to right). The azimuthal position of the planet is $\phi=225^{\circ}$. The beam resolution is 40 mas.
    }
    \label{fig:channel_DIFF_vres}
\end{figure*}

Fig.~\ref{fig:channel_DIFF_vres} shows channel maps obtained by changing the channel velocity resolution in the range [0.1,0.2,0.3,0.5] km/s (from left to right). We kept the planet fixed at an azimuthal location of $\phi=225^{\circ}$. The beam size is set to 40 mas. The kink feature is detectable with all the velocity resolutions.

To summarise, with this more detailed analysis in which we vary both the beam and velocity resolution we found that a planet kink in this system would be detectable using a beam of [30,40] mas (barely visible with 50 mas), and velocity resolution between 0.1 and 0.5 km/s. A more detailed study about the probability of detecting or not a planet depending on its azimuthal position would be necessary. Moreover, also if a kink is not detectable, having access to gas kinematics data could be useful to map the height of the two side of the CO-isotopologue layer in the disc and check if there are some discrepancies due to the presence of a planet \citep{pinte18}.

\section{Conclusions}\label{sec:conclusion}
In this paper we modeled the gap shape observed by \cite{long18} and Long et al. (2020, submitted) around the DS Tau star. Assuming that a planet has carved the gap, we performed 3D dust and gas Smoothed Particle Hydrodynamics and radiative transfer simulations of protoplanetary discs with one embedded planet, considering different values for the planet mass.By comparing the simulated dust gap/ring morphology with observations, we derive a best-fit planet mass and compare the result with simple analytical calculations \citep{lodato19}. We also studied the gas kinematics in order to check if a kink produced by the embedded planet would be visible in the channel maps. 

The basic result of our modeling is that the planet we expect to be responsible for the gap observed in DS Tau should have a mass of $M_{\rm p}=3.5\pm1\,M_{\rm Jup}$. We recall that this confidence interval should be interpreted in the light of the chosen one-planet model. For future developments, it would be interesting to gradually add other degrees of freedom in order to explore other formation scenarios.
To reach this result, we performed a $\chi^2$ test comparing the major axis radial profile of the data with the ones obtained by the different planet mass models. To match the correct flux, we also performed different radiative transfer models varying the gas-to-dust ratio from our initial reference case (i.e. 100) in order to take into account different dust disc masses. The outcome of this fitting procedure is that the dust mass required to produce the observed ALMA images is of $9.6\cdot10^{-5}\,M_{\odot}$. Another interesting point to be noted is that in order to recover the correct gap shape, it is necessary to study the system in its time evolution. Indeed, as time increases, the inner disc is gradually depleted, paving the way for a cavity to originate at a later stage (see Sec.~\ref{sec:fit_continuum} for a discussion about possible numerical effects that might affect the dust depletion in the inner disc). 

Starting from this result, we then computed the CO,$^{13}$CO and C$^{18}$O isotopologue channel maps. We found that a planet with 3.5 Jupiter masses, with a channel width of 0.3/0.5 km/s and a beam size of 70 mas, would be barely detectable in the gas kinematics through the kink feature. Instead, by assuming a slightly higher (0.2 km/s) velocity resolution, and by choosing a slightly smaller (i.e. 40 or 50 mas) beam size the kink appears to be visible. Moreover, we showed that changing the azimuthal position of the planet results in different kink signatures.

\section*{Acknowledgements}

We thank the referee Ruobing Dong for an insightful report of the manuscript. We thank the SPHgroup at the Monash University in Melbourne for fruitful discussions that improved the manuscript.
GL and BV have received funding from the European Union’s Horizon 2020 research and innovation programme under the Marie Skłodowska-Curie grant agreement No 823823 (RISE DUSTBUSTERS project).
ER acknowledges financial support from the European Research Council (ERC) under the European Union's Horizon 2020 research and innovation programme (grant agreement No 681601). DP, CP and VC acknowledge funding from the Australian Research Council via FT130100034 and DP180104235.

We used OzStar, funded by Swinburne University of Technology and the Australian government. We used \textsc{phantom} \citep{price18phantom} for the hydrodynamic simulations and \textsc{mcfost} \citep{pinte06,pinte09} for the radiative trasnfer calculations.
We used \textsc{splash} \citep{splash} for rendered images of our simulated hydrodynamic systems, the \textsc{pymcfost} tool provided by Christophe Pinte for the rendering of the channel maps, while the remaining figures have been generated using the \textsc{python}-based MATPLOTLIB package  \citep{matplotlib}.




\bibliographystyle{mnras}
\bibliography{biblio} 



\appendix

\section{CO-isotopologues channel maps}\label{app:channels}
\subsection{Channels for $v_{\rm res}=0.2$ km/s and with beam resolution 40 mas}
\begin{figure*}
 	\includegraphics[scale=0.3]{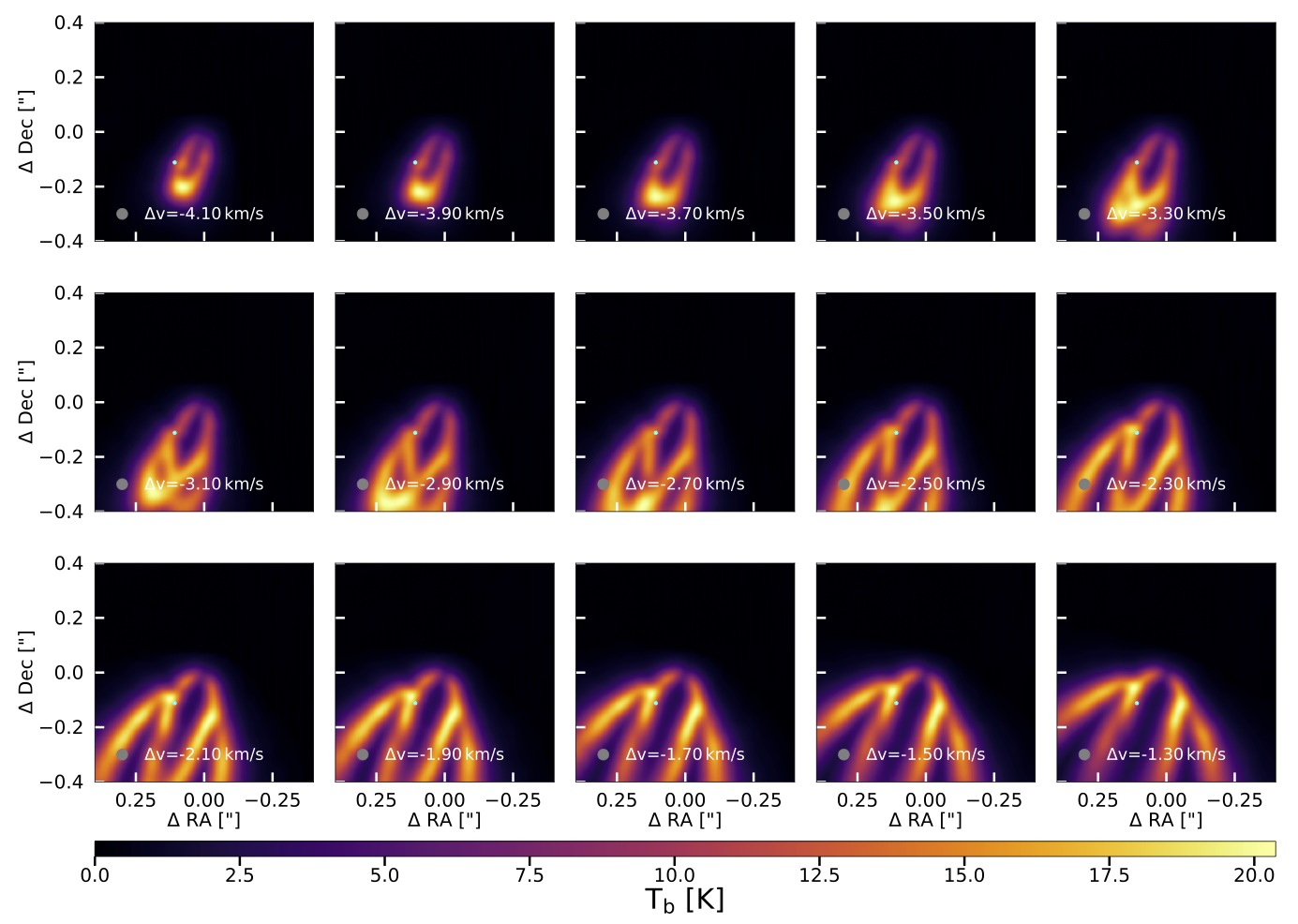} 
 	\caption{Channel maps for C$^{18}$O isotopologue with a velocty resolution of 0.2 km/s, from -4.10 to -1.30 km/s. The azimuthal position of the planet is $\phi=225^{\circ}$. The beam resolution is 40 mas. 
    }
    \label{fig:ALLchannels}
\end{figure*}

\begin{figure*}
 	\includegraphics[scale=0.3]{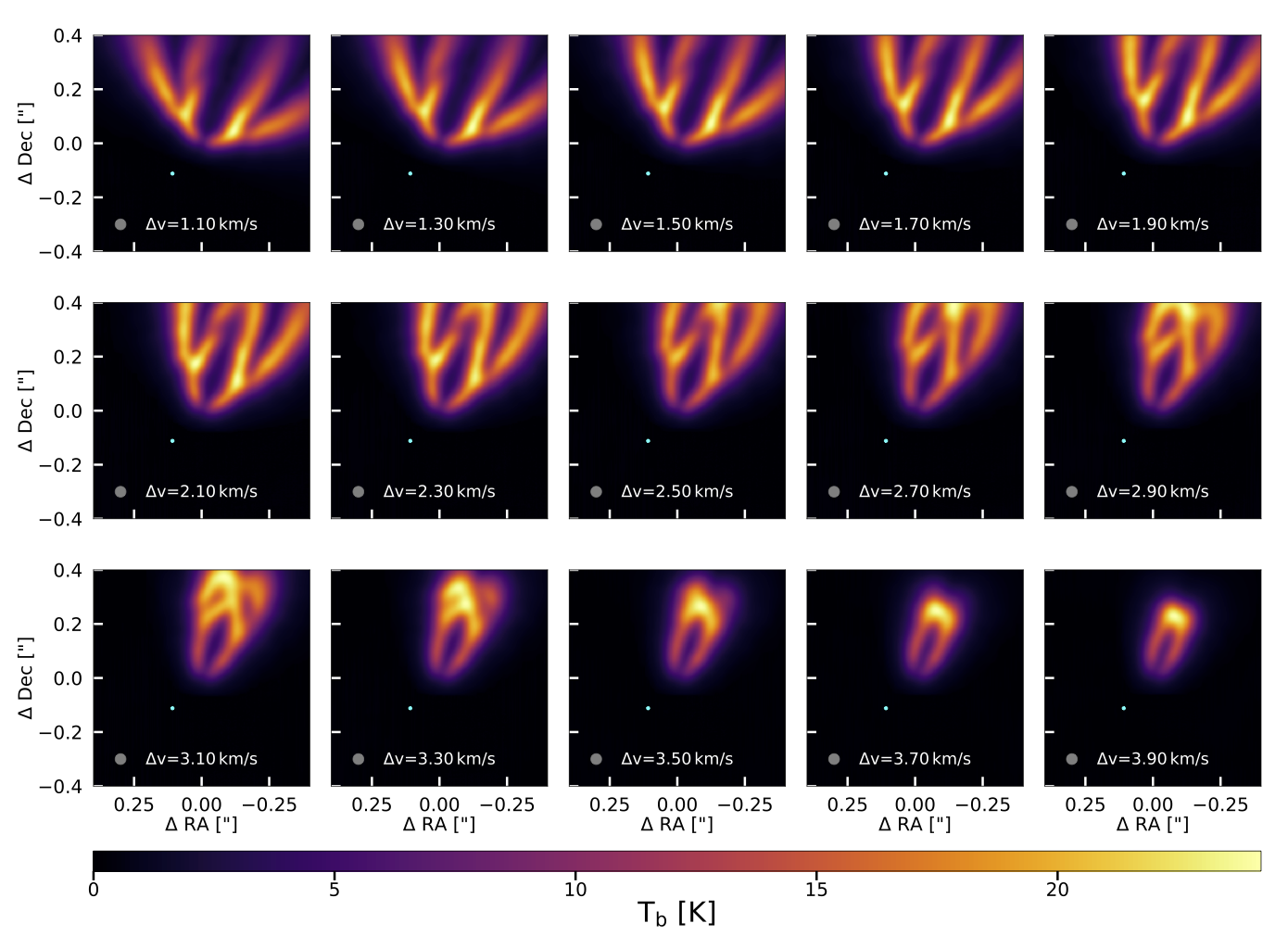}
 	\caption{Channel maps for C$^{18}$O isotopologue with a velocity resolution of 0.2 km/s, from 1.30 to 4.10 km/s. The azimuthal position of the planet is $\phi=225^{\circ}$. The beam resolution is 40 mas. 
    }
    \label{fig:ALLchannels2}
\end{figure*}

Fig.~\ref{fig:ALLchannels} and Fig.~\ref{fig:ALLchannels2} shows velocity channels for the model with a 3.5 $M_{\rm Jup}$ planet (represented with the outer cyan dot). The velocity resolution is 0.2 km/s, for channels going from -4.10 to -1.30 km/s and from 1.30 to 4.10 km/s. The beam resolution is 40 mas. 

\subsection{Minimum detectable planet mass test: $M_{\rm p}=2.3\,M_{\rm Jup}$ }\label{app:1mjup}
\begin{figure*}
 	\includegraphics[scale=0.45]{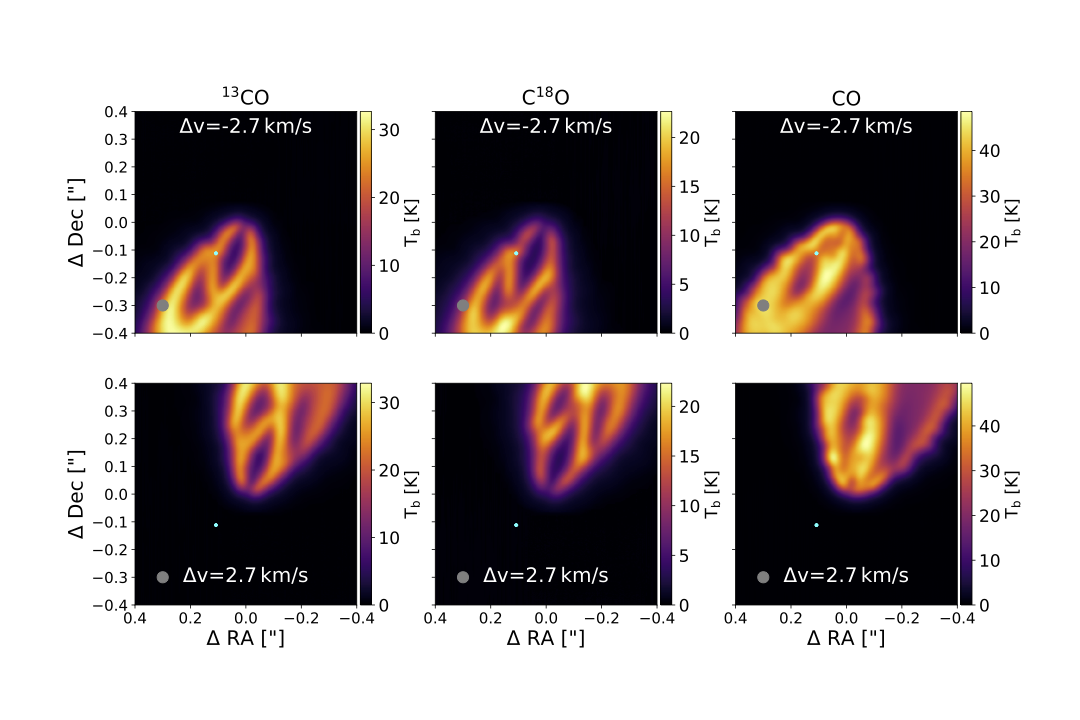}
 	\caption{Channel maps obtained for a planet with mass $M_{\rm p}=2.3\,M_{\rm Jup}$. Columns show the $^{13}$CO (left column), C$^{18}$O (center column), CO (right column) isotopologue channel maps with a velocity resolution of 0.2 km/s, $\phi=225^{\circ}$ and with a beam resolution of 40 mas. The velocity channel of the first row is the one in which the kink is visible ($\Delta v=-2.7$ km/s), while the second row shows its opposite ($\Delta v=2.7$ km/s). 
    }
    \label{fig:chann_2.5Mjup}
\end{figure*}

We present here the channel maps computed for a planet with a mass of $2.3\,M_{\rm Jup}$. We perform this test to check what is the minimum planet mass detectable by ALMA at the distance of $\approx 30$ au from the star, according to our models. As reference case, we use the best observational parameters we found in Sec.~\ref{sec:kink}: an angular resolution of 40 mas, an azimuthal planet position of $\phi=225^{\circ}$ and a velocity resolution of $v_{\rm res}=0.2\rm{kms}^{-1}$. Fig.~\ref{fig:chann_2.5Mjup} shows the three CO-isotopologue channel map (CO left column,$^{13}$CO center column ,C$^{18}$O right column), for two different channels, with $\Delta$v equals to -2.7 (top row) and 2.7 (bottom row) km/s from the systemic velocity. The channel maps we recover for this planet are very similar to the one presented in Sec.~\ref{sec:kink}. The kink should be detectable also in this case in the channel with $\Delta v=-2.7$ km/s. 


\bsp	
\label{lastpage}
\end{document}